\documentclass[aps,prl,twocolumn,float]{revtex4}
\usepackage{amsmath,bm,epsfig}
\newcommand{\B}[1]{{\bm{#1}}}%% Bold Roman & Greek Lower & Upper Case
\usepackage[dvips]{color}
\begin{document}
%%%%%%%%%%%%%%%%%%%%%%

\title{Scaling Theory of the Mechanical Properties of Amorphous Nano-Films}
\author{Awadhesh K. Dubey$^1$, H. George E. Hentschel$^2$, Prabhat K.
Jaiswal$^1$, Chandana Mondal$^1$, Yoav G. Pollack$^1$, and Itamar Procaccia$^1$}
\affiliation{$^1$Department of Chemical Physics, The Weizmann Institute of
Science,  Rehovot 76100, Israel\\$^2$ Department of Physics,
Emory University, Atlanta, Georgia}
\begin{abstract}
Numerical Simulations are employed to create amorphous nano-films of a chosen
thickness on a crystalline substrate
which induces strain on the film. The films are grown by a vapor deposition
technique which was recently developed
to create very stable glassy films. Using the exact relations between the
Hessian matrix and the shear and bulk
moduli we explore the mechanical properties of the nano-films as a function of
the
density of the substrate and the film thickness. The existence of the substrate
dominates the
mechanical properties of the combined substrate-film system. Scaling concepts
are then employed to achieve data collapse in a wide range of densities and film
thicknesses.
\end{abstract}
\maketitle
%%%%%%%%%%%%%%%%%

Numerical simulations are notorious for their innate inability to explore large
systems of experimental
relevance. This shortcoming can be turned into an advantage in the study of
nano-systems, providing
a precious system-size dependence of the explored properties. Indeed,
nanometer-thick  films  with
designed compositions  and microstructures occupy a central position in
nanomaterials research.

Thin films are everywhere around us and we are very much dependent on them in
our everyday life.
Thin films find huge applications in science and technology
and their importance keeps on growing for a wide range of applicabilities
\cite{thinfilm,semiconductor,led,optical-coating} in
areas such as electronic semiconductor devices, LEDs, optical coatings,
hard coatings on cutting tools, for both energy,
generation, e.g., thin film solar cells and storage. Amorphous thin films
composed of rare-earth transition metals are heavily used for producing
magnetic data storage devices \cite{application1,application2}.

Though there have been plenty of experimental studies on both crystalline and
amorphous thin films, there are very few theoretical and simulation studies
\cite{theory1,theory3}. In particular a systematic investigation of the
film-thickness and substrate density dependence of mechanical properties of
amorphous thin films is still lacking despite their obvious technological
importance. In most of the studies the main focus is bestowed upon the
electronic and magnetic properties \cite{mag1,mag2,mag3}. However, it is equally
important to investigate their  mechanical properties as these devices must be
reliable, they must have structural integrity,
and they must retain that integrity over their lifetime, mechanical failures
must not occur.

In this Letter we explore how numerical simulations can shed new light on the
mechanical properties of amorphous films grown on crystalline substrates. The
central point of this Letter will be that using scaling concepts one can provide
a predictive theory for the changes in the mechanical properties when the
density of the substrate and the thickness of the film vary.

We will limit our study in this paper to the mechanical properties of thin films
on substrates as the substrate plays a key role in the properties of thin films.
The substrate can induce huge stresses
\cite{substrate-stress1,substrate-stress3,substrate-stress4} on the film the
nature of which depends on various factors such as the geometry of the
substrate, lattice mismatch and difference in thermal expansion coefficients of
the film and substrate etc.

The model discussed in the Letter grows a film of a binary glass on top of
a crystalline substrate.
The  binary glass film is made from 65\% particles A and 35\% particles B,
interacting via Lennard Jones (LJ) potentials with parameters
$\lambda_{BB}/\lambda_{AA}=0.88$, $\lambda_{AB}/\lambda_{AA}=0.8$,
$\epsilon_{BB}/\epsilon_{AA}=0.5$ and $\epsilon_{AB}/\epsilon_{AA}=1.5$. This
system has been extensively studied as a model glass former~\cite{KA1,KA2,KA3}.
Lengths and energies are given in terms of $\sigma_{AA}$ and $\epsilon_{AA}$,
while time units are given by $\sqrt{ m \lambda_{AA}^{2}/\epsilon_{AA}}$.  Both
the Boltzmann constant $k_B$ and the mass of the particles are taken to be unity
throughout. All the pair interactions are truncated at distance $r_c$ where
$r_c/\lambda=2.5$ with two continuous derivatives.

 The films are grown in our simulations by a vapor deposition technique
\cite{vap-dep1,vap-dep2} which closely mimics the physical vapor deposition
technique in experiments where thin films are prepared by depositing hot molecules onto a
substrate. In our simulations, we generated a crystalline
substrate having four atomic layers along the $z$ direction composed of
$N_S=24\times 28\times 4$ particles interacting via Lennard-Jones with the
characteristic size and energy: $\lambda_{SS}/\lambda_{AA}=0.8$,
$\lambda_{SA}/\lambda_{AA}=0.8$, $\lambda_{SB}/\lambda_{AA}=0.8$,
$\epsilon_{SS}/\epsilon_{AA}=1.5$, $\epsilon_{SA}/\epsilon_{AA}=1.5$ and
$\epsilon_{SB}/\epsilon_{AA}=1.5$. The substrate particles were organized in an
hcp planar arrangement, with atoms restrained to their positions by a harmonic
potential with spring constant $K = 1000$. The simulation box was kept periodic
only in the $x$ and $y$ directions and the $x$ and $y$ box lengths are determined
by the density $\rho_{S}$ of the substrate. The volume of the system was kept
constant, but the length of the simulation box in the $z$ direction
(perpendicular to the substrate) was sufficiently large to encompass a growing
glass film and a vacuum region into which molecules were gradually introduced.

\begin{figure}
\includegraphics[scale = 0.47]{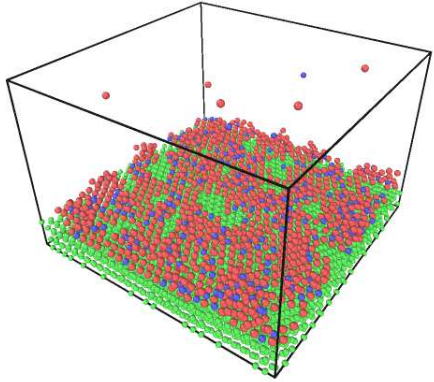}
\caption{Visualization of the process of deposition on the crystalline substrate
of an amorphous
film. In green are the substrate particles, and red and blue are particles A and
B, respectively.}
\label{visual}
\end{figure}
The actual algorithm consisted of depositions of a layer after layer of $A$ and
$B$ particles, about $100$ at a time, very close to the top of the free-surface.
Each such deposition was followed by a procedure where the substrate and any
previously deposited particles were maintained at the temperature $T_S=0.1$,
while simultaneously the newly added particles were first equilibrated at a hot
temperature $T_H = 0.7$, then cooled down to $T_S$ at a rate of
$3.33\times10^{-2}$, and lastly equilibrated at $T_S$ through the use of two
Nos\'e-Hoover chains ~\cite{vap-dep1,vap-dep2} thermostat. We choose $T_S=0.1$
such that it is well below the reported glass transition temperature ($T_g$) of
the Kob-Anderson $65:35$ binary mixture in $3D$ ~\cite{KA6535Tg}.
The process of deposition on top of the substrate is visualized in
Fig~\ref{visual}.

After a prescribed number of particles have been deposited, depositions stopped
and measurements are carried out after simulating the whole system at $T_s$ for
$120000$ time units in steps of $\delta t=0.01$. Measurements were performed
$100$ times for a total period of $2000$ time units. We define two variables,
the deposited film thickness $w_D$ (according to substrate configuration) and
the actual film thickness $w$. The number of particles to be deposited is
determined, at the beginning of a simulation, by the formula $N_D=Lx\times
Ly\times w_D\times 1.2$, where $w_D$ is varied from $0.5$ to $10.0$. However,
after deposition the top surface is not smooth and $w$ is calculated by dividing
the $xy$-area in small blocks and calculating the local height and averaging it
over the blocks.

%%%%%%%%%%%%%%%%%%%%%%%%%%%%%%%%%%%%%%%%%%%%%%%%
\begin{figure}
\includegraphics[scale = 0.47]{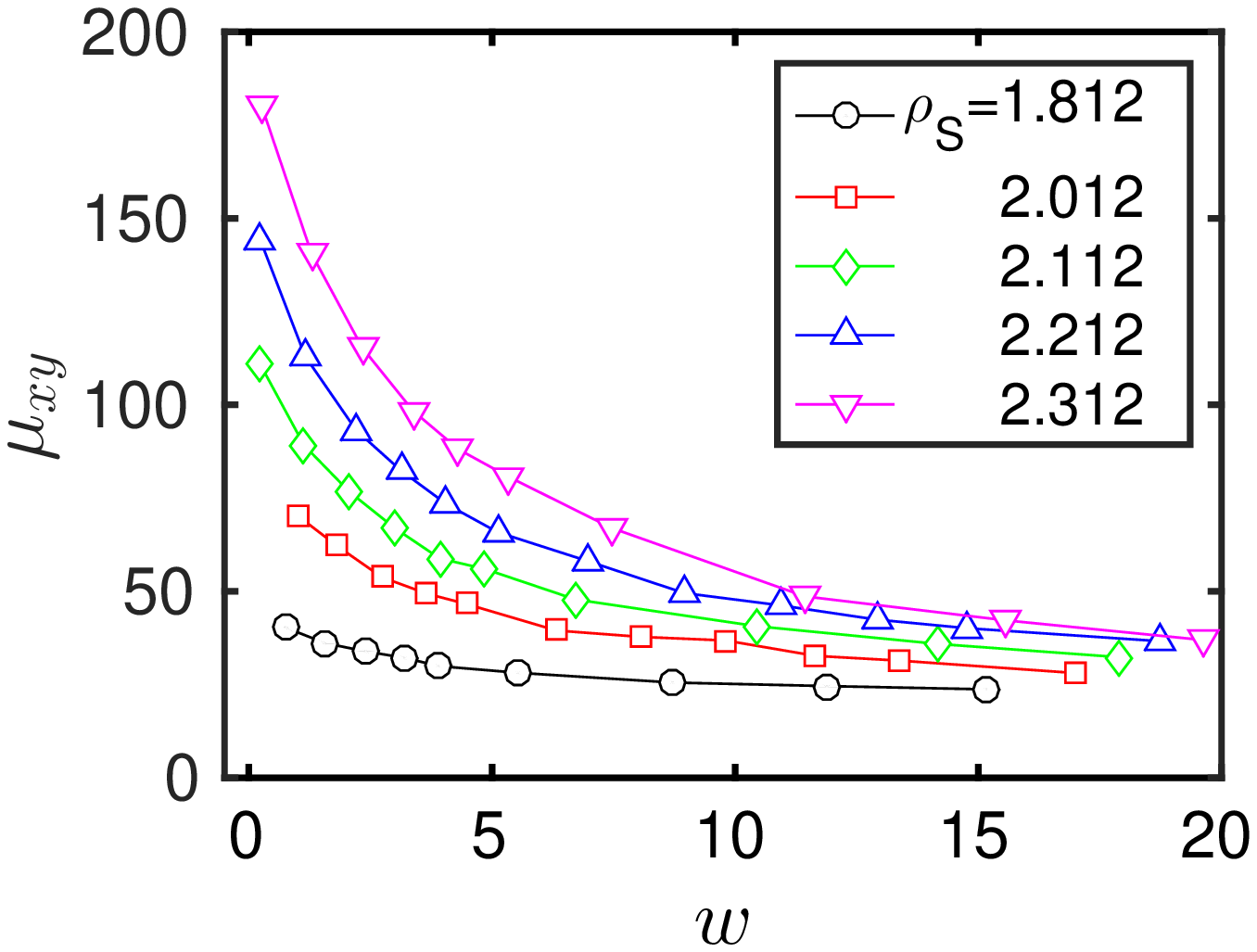}
\includegraphics[scale = 0.47]{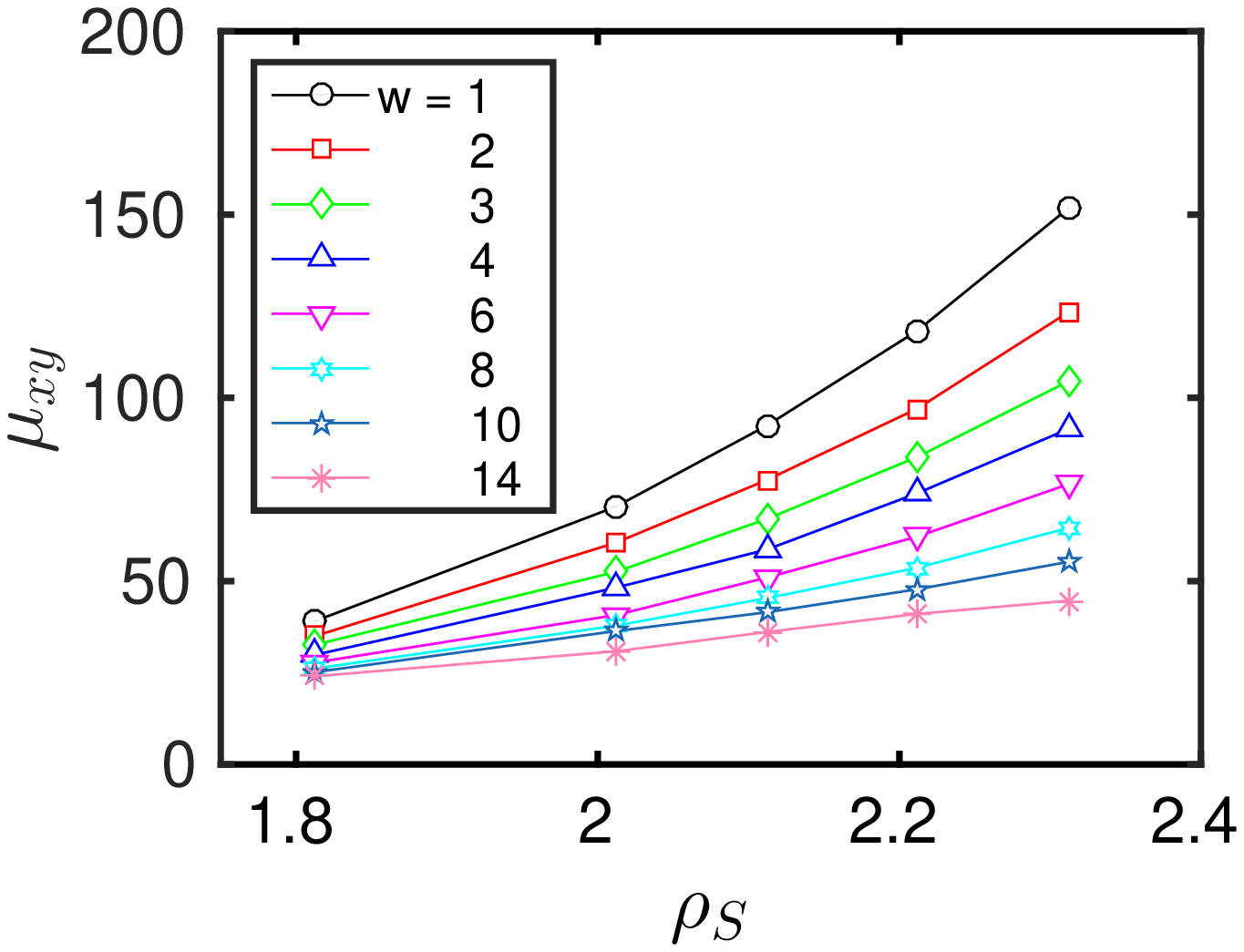}
\caption{Upper panel: the $xy$ shear modulus as a function of film width for
different values of the
substrate density. Lower panel: the same modulus as a function of substrate
density for different
film widths.}
\label{shearmod}
\end{figure}
%%%%%%%%%%%%%%%%%%%%%%%%%%%%%%%%%%%%%%%%%%%%%%%%%%%%%%
Having constructed the film of desired thickness over the substrate, we can
compute its mechanical properties.
Thus for example the shear modulus $\mu_{xy}$ is obtained exactly as
\begin{equation}
\mu_{xy} = \mu_{xy}^B -\frac{V}{T} [\langle \sigma_{xy}^2\rangle -\langle
\sigma_{xy}\rangle^2] \ ,
\end{equation}
where $\sigma_{xy}$ is the $xy$ component of the fluctuating stress tensor
\cite{fluctuation} and $\mu_{xy}^B$ is the Born
term \cite{born} which has the explicit form
\begin{eqnarray}
\mu^B_{xy} &=& \frac{1}{V}\sum_{j>i} \left[y_{ij}^2\frac{1}{r_{ij}}\frac{\partial
U}{\partial r_{ij}} + x_{ij}^2y_{ij}^2\left(\frac{1}{r_{ij}^2}\frac{\partial^2
U}{\partial r_{ij}^2} - \frac{1}{r_{ij}^3}\frac{\partial U}{\partial
r_{ij}}\right)\right] \nonumber \\ &+& K\sum_{i=1}^{N_S}(y_i-y_{i_0})^2 ,
\label{muxyb}
\end{eqnarray}
where $y_{i_0}$ is the reference $y$-coordinate of the $i^{th}$ substrate-particle.
The molecular dynamics code can be used now to measure the fluctuation in the
stress tensor as well as the
fluctuating inter-particle distances $\B r_{ij}$. Using this information both
terms in the shear modulus
can be computed. The computation is repeated twenty times and the shear modulus
is averaged over these twenty runs.  The results are shown in
Fig.~\ref{shearmod} as a function of the film width $w$ and of the
substrate density $\rho_S$.
Note the dramatic effect of the substrate density on the shear modulus, changing
it (for small film width)
by a factor of four when the substrate density changes by 25\%. Clearly, this
strong effect disappears
upon increasing the film thickness and asymptotically the shear modulus
converges to the bulk limit.

%%%%%%%%%%%%%%%%%%%%%%%%%%%%%%%%%%%%%%%%%%%%%%%%
\begin{figure}
\includegraphics[scale = 0.47]{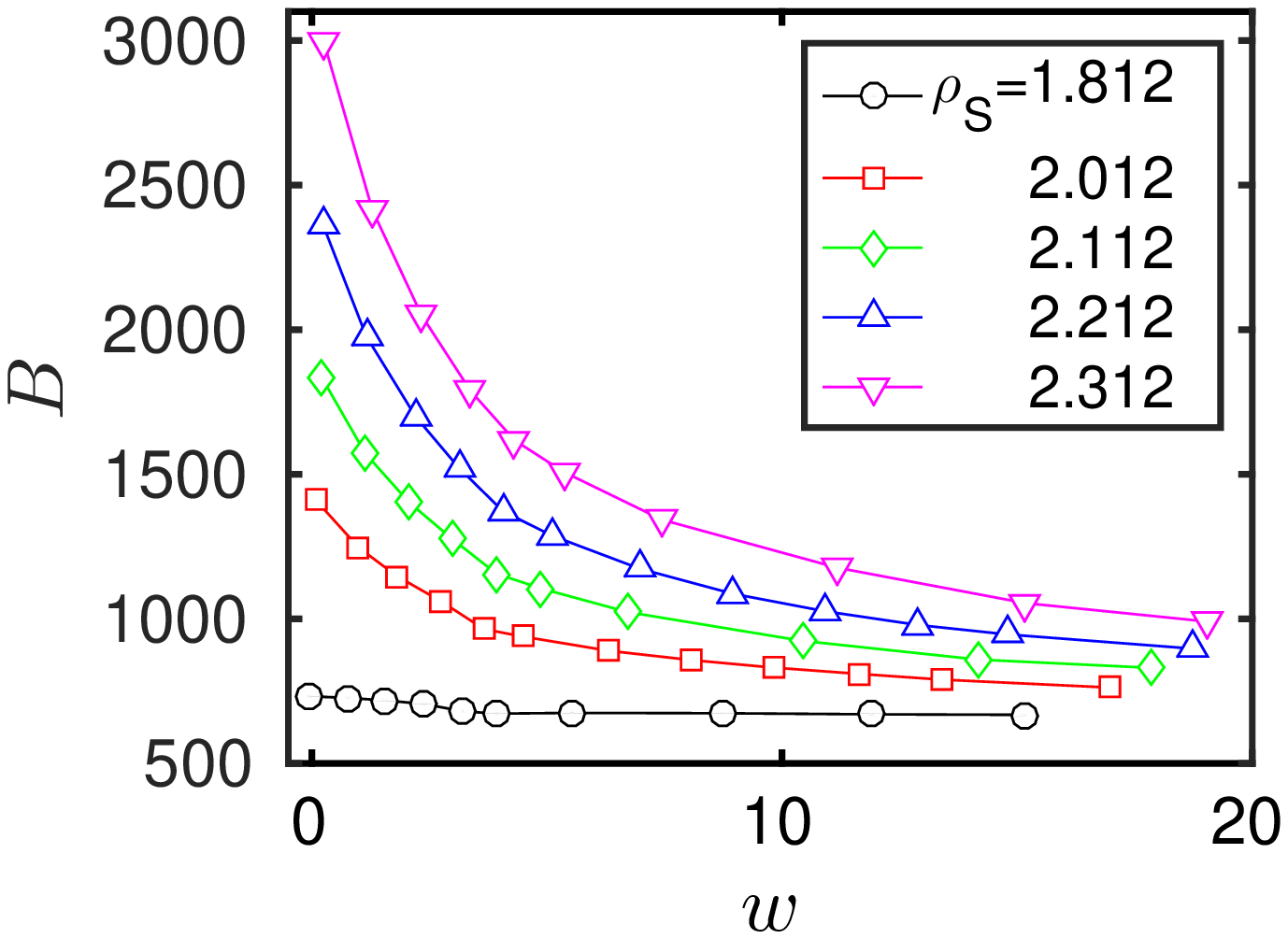}
\includegraphics[scale = 0.47]{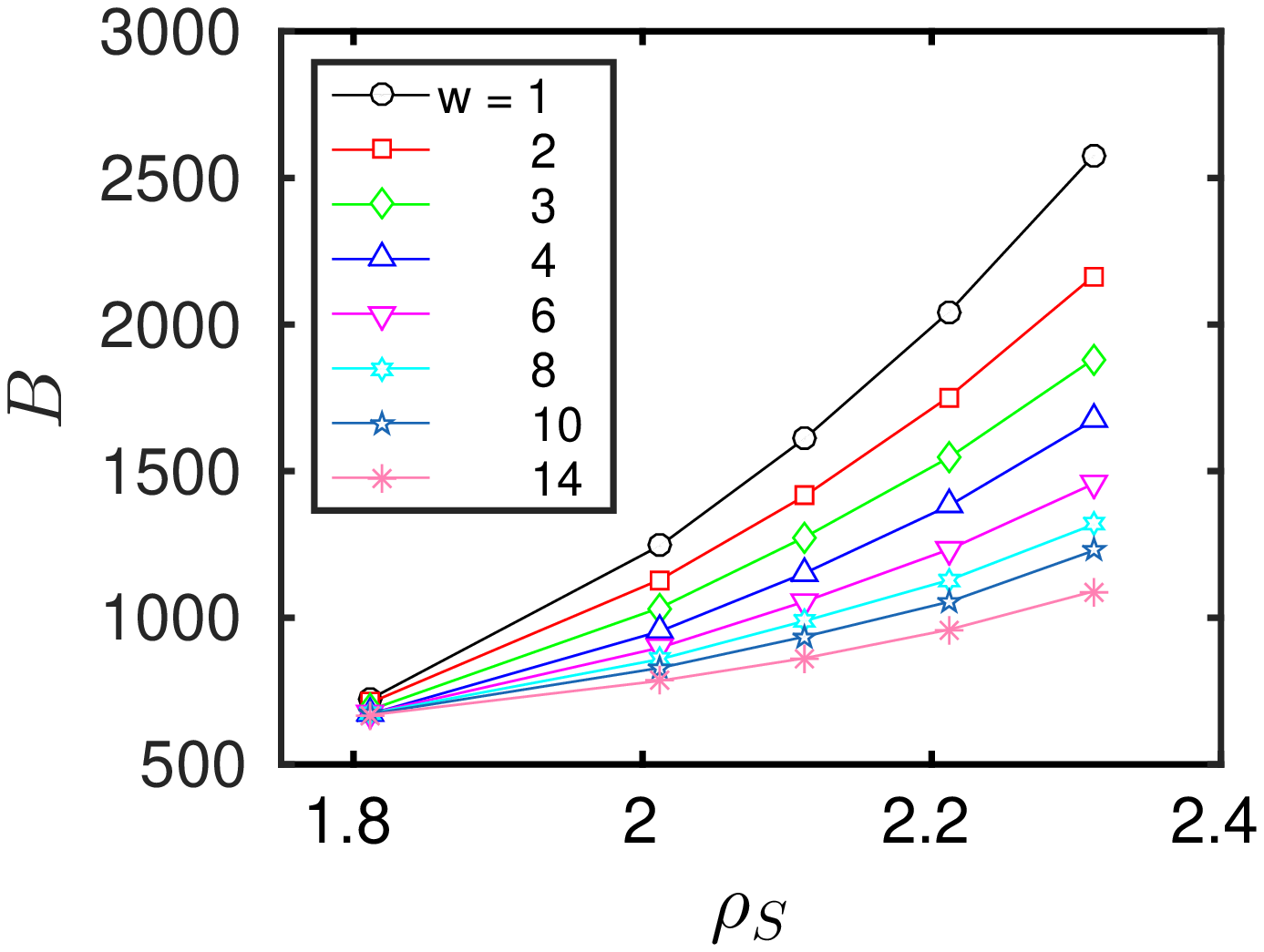}
\caption{Upper panel: the bulk modulus as a function of film width for different
values of the
substrate density. Lower panel: the same modulus as a function of substrate
density for different
film widths.}
\label{bulkmod}
\end{figure}
%%%%%%%%%%%%%%%%%%%%%%%%%%%%%%%%%%%%%%%%%%%%%%%%
Similarly to the shear modulus the bulk modulus is given by the exact expression
\begin{equation}
B = B^B -\frac{V}{T} [\langle P^2\rangle -\langle P\rangle^2] \ ,
\end{equation}
where $P$ is the fluctuating pressure and $B^B$ is the Born
term which has the explicit form
\begin{eqnarray}
B^B &&=\frac{1}{V}\sum_{j>i}
\left[(x_{ij}^2+y_{ij}^2+z_{ij}^2)\frac{1}{r_{ij}}\frac{\partial U}{\partial r_{ij}} \right.
\nonumber \\
&&+ \left. (x_{ij}^2+y_{ij}^2+z_{ij}^2)^2\left(\frac{1}{r_{ij}^2}\frac{\partial^2
U}{\partial r_{ij}^2} - \frac{1}{r_{ij}^3}\frac{\partial U}{\partial r_{ij}}\right)\right]
\nonumber \\ &&+ K\sum_{i=1}^{N_S}\left[(x_i-x_{i_0})^2+(y_i-y_{i_0})^2+(z_i-z_{i_0})^2\right].
\label{Bb}
\end{eqnarray}
Here, $x_{i_0}$ and $z_{i_0}$ are the reference $x$- and $z$-coordinates of the $i^{th}$ substrate-particle.
Both terms are readily computed in the MD code, and the average computed from
twenty independent runs are shown in Fig.~\ref{bulkmod}.

\begin{figure}
\includegraphics[scale = 0.47]{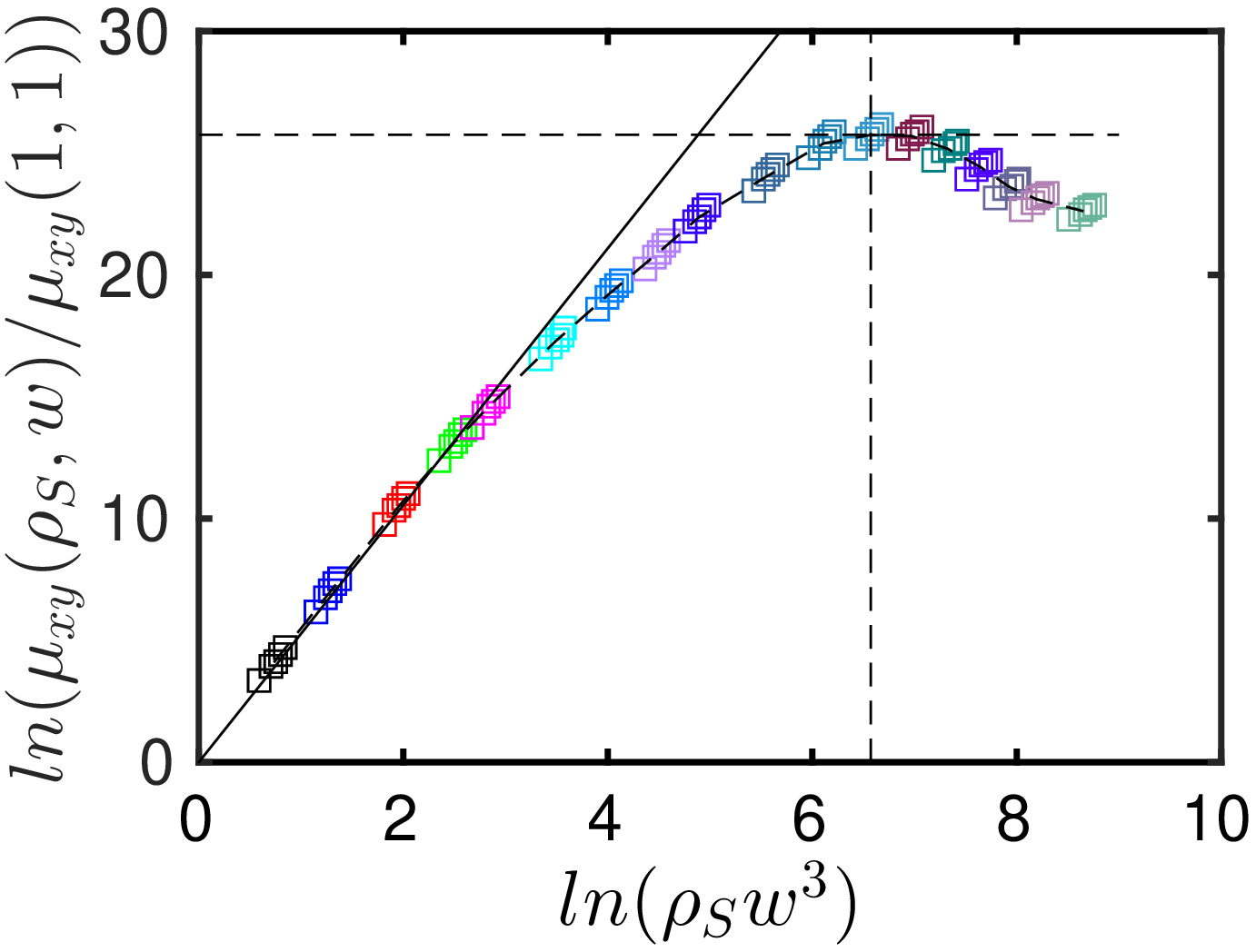}
\includegraphics[scale = 0.47]{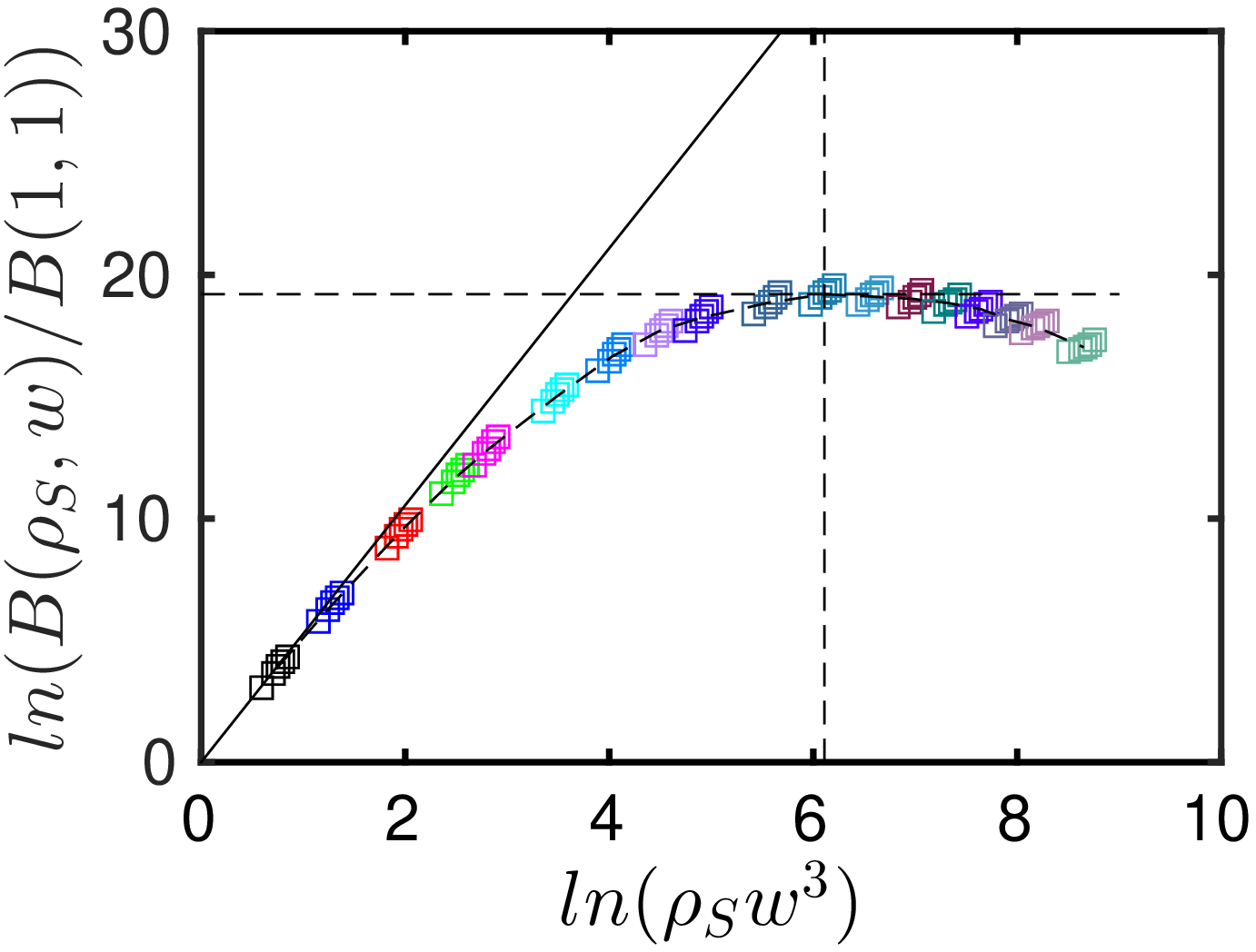}
\caption{The data collapse for the shear modulus (upper panel) and for the bulk
modulus (lower panel).}
\label{datacol}
\end{figure}

The effect of the substrate is as dramatic for the bulk modulus as for the shear
modulus. Our next task
is to understand these results and turn them into a predictive theory.
Both the shear and the bulk moduli are functions of two variables, i.e.,
$\rho_S$ and $w$. It is advantageous
to represent both objects as a function of a dimensionless variable, i.e.,
$x=\rho_S w^3$. Needless to say, we
need to preserve dimensions, so we try a representation in the form
\begin{eqnarray}
\mu_{xy}(\rho_S,w) = \mu_{xy}(1,1) f_\mu (\rho_Sw^3)\ , \\
B(\rho_S,w) = B(1,1) f_B (\rho_Sw^3) \ .
\end{eqnarray}
The form of the scaling functions $f_\mu$ and $f_B$ can be gleaned from the data
collapse shown in
Fig.~\ref{datacol}.
%%%%%%%%%%%%%%%%%%%%%%%%%%%%%%%%%%%%%%%%%%%%%%%%
%%%%%%%%%%%%%%%%%%%%%%%%%%%%%%%%%%%%%%%%%%%%%%%%%%%%
The almost perfect data collapse begs the next questions: (i) what is the slope
of the functions
$f_\mu(x)$ and $f_B(x)$ in the limit $x\to 1$ and (ii) what determines the
position of the maximum
of these two functions. Obviously, these must be determined by the parameters of
the model.

\begin{figure}
\includegraphics[scale = 0.47]{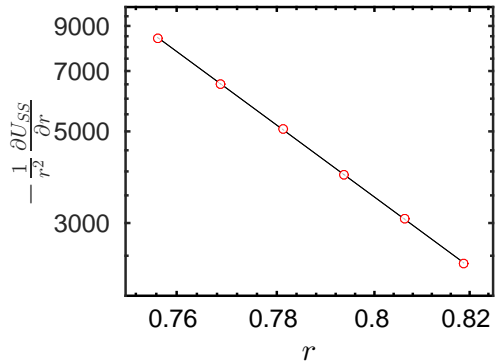}
\caption{The effective power law in the range of scales determined by the range
of substrate densities $r_0/\lambda\in [\rho_{S,max}^{-1/3},\rho_{S,min}^{-1/3}]$.}
\label{effective}
\end{figure}

\begin{figure}
\includegraphics[scale = 0.47]{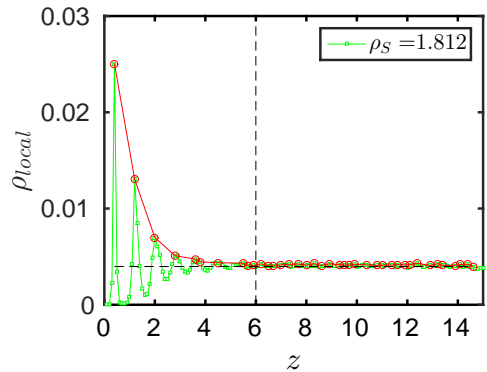}
\caption{The local density of the glassy film as a function of the $z$-coordinate (green).
The influence length of the substrate is estimated from here to be $\xi_S\approx
6$. The envelope of the data is shown in red color.}
\label{locdens}
\end{figure}

To understand the slope we examine the potential of interaction between the
substrate particles. This interaction defines a typical scale
\begin{equation}
r_0\sim \frac{\lambda_{SS}}{\rho^{1/3}_{S}} \ .
\end{equation}
In Fig.~\ref{effective} we demonstrate that in the range of substrate densities
considered here
the interaction is represented to a very good approximation as an effective
power law \cite{09LP}
\begin{equation}
-\frac{1}{r^2}\frac{\partial U_{SS}(r)}{\partial r} \sim
\frac{\epsilon_{SS}}{\lambda_{SS}^3} \left(\frac{r}{\lambda_{SS}}\right)^{-3\nu}
\ ,
\label{defnu}
\end{equation}
which serves as a definition of the effective exponent $\nu$ which in our case
takes on the value $\nu=5.27$.
%%%%%%%%%%%%%%%%%%%%%%%%%%%%%%%%%%%%%%%%%%%%%%%%%%%%%%%%%%%%%%%%
%%%%%%%%%%%%%%%%%%%%%%%%%%%%%%%%%%%%%%%%%%%%%%%%%%%%%%%%%%%%%%%%
In the limit $w\to 1$ the shear modulus is dominated by the substrate density,
and from dimensional
considerations we expect that \cite{09LP}
\begin{equation}
\mu_{xy}(x\to 1) \sim \frac{\epsilon_{SS}}{\lambda_{SS}^3}
\rho_S^\nu \ .
\label{murho}
\end{equation}
Indeed the initial slope in Fig.~\ref{datacol} agrees very well with this
estimate. The argument is
identical for the bulk modulus (the same dimensional considerations) and indeed
the initial slopes
are the same for both quantities. We thus end up with the prediction
\begin{equation}
f_\mu(x)\sim f_B(x) \sim x^\nu \ , \quad \text{for}~x\ll x_c \ ,
\label{smallx}
\end{equation}
where $\ln{x_c}$ is the point of maximum in Fig.~\ref{datacol}. The other limit is
obvious, reading as $x\to \infty$
\begin{equation}
f_\mu(x) \to \frac{\mu_{xy}^\infty}{\mu_{xy}(1,1)} \ , \quad f_B(x)
\to \frac{B^\infty}{B(1,1)} \ .
\end{equation}
where $\mu_{xy}^\infty$ and $B^\infty$ are the shear and bulk moduli for the
pure glassy phase. Finally we
need to understand the point of maximum $\ln x_c$. This must be determined by the
typical scale, say  $\xi_S$, characterizing
the influence of the substrate on the properties of the film. This length can be
read straightforwardly from
the local density of the film as a function of the coordinate $z$. A typical
such plot is shown in Fig.~\ref{locdens},
%%%%%%%%%%%%%%%%%%%%%%%%%%%%%%%%%%%%%%%%
from which we can estimate $\xi_S\approx 6$. Since $\rho_S \approx 2$, we can
estimate $\ln x_c\approx \ln(\rho_S\xi_S^3)\approx 6$ as seen indeed in
Fig.~\ref{datacol}.

Having at hand  scaling functions one can predict the values of the shear and
bulk moduli for any value of
$w$ or $\rho_S$ where these properties are not measured. Moreover, any quantity
with the same dimensions
is expected to have a scaling function having the same initial slope and a
maximum at the same value of $w^3\rho_S$.
Other mechanical properties like the Young modulus and the Poisson ratio also
succumb to scaling concepts, but
the presentation of that theory is beyond the scope of the present Letter and
will be expounded in a
forthcoming publication.

\acknowledgments
This work had been supported in part by an ``ideas" grant STANPAS of the European
Research Council (ERC) and by the Minerva
Foundation, Munich, Germany. We acknowledge with thanks fruitful discussions
with Shiladitya
Sengupta and Murari Singh. PKJ acknowledges VATAT fellowship from the Council of Higher Education, Israel.

%%%%%%%%%%%%%%%%%%%%%%%%%%%%%%%%%%%%%%%%%%%

\end{document}